\documentclass[cits]{PoS}
\pdfoutput=1

\usepackage{microtype}
\usepackage{amsmath}
\usepackage{subfig}

\newcommand{\pbp}{\langle\bar\psi\psi\rangle}
\newcommand{\dmin}{d_{\text{min}}}

\DeclareMathOperator{\tr}{tr}

\title{Scalar correlators near the 3-flavor thermal critical point}

\ShortTitle{Scalar correlators near the 3-flavor thermal critical point}

\author{
  \speaker{Xiao-Yong Jin}$^{\dag a}$, 
  Yoshinobu Kuramashi$^{abc}$,
  Yoshifumi Nakamura$^{ad}$,
  Shinji Takeda$^{ae}$,
  and
  Akira Ukawa$^a$
  \vspace{.3em}\\
  \llap{$^a$}RIKEN Advanced Institute for Computational Science, Kobe, Hyogo 650-0047, Japan
  \vspace{.3em}\\
  \llap{$^b$}Graduate School of Pure and Applied Sciences, University of Tsukuba, Tsukuba, Ibaraki 305-8571, Japan
  \vspace{.3em}\\
  \llap{$^c$}Center for Computational Sciences, University of Tsukuba, Tsukuba, Ibaraki 305-8577, Japan
  \vspace{.3em}\\
  \llap{$^d$}Graduate School of System Informatics, Department of Computational Sciences, Kobe University, Kobe, Hyogo 657-8501, Japan
  \vspace{.3em}\\
  \llap{$^e$}Institute of Physics, Kanazawa University, Kanazawa 920-1192, Japan
  \vspace{.3em}\\
  \llap{$^\dag$}Email: \email{xjin@anl.gov}\\
  Present address: Argonne Leadership Computing Facility, Argonne National Laboratory, Argonne, Illinois 60439, USA
}

\abstract{We investigate screening masses at both sides of the first
  order finite temperature transition with 3 quark flavors using the
  nonperturbatively improved clover fermion action and the Iwasaki
  gauge action.  We have developed the method of hierarchical
  truncations with stochastic probing to accelerate the noise
  estimator for evaluating quark loops at every spatial lattice
  slices.  At parameter values we study, the flavor singlet scalar
  meson has a screening mass about half of the pion screening mass.
  It becomes lighter as the system approaches the critical endpoint.}

\FullConference{The 32nd International Symposium on Lattice Field Theory\\
                 23-28 June, 2014\\
                 Columbia University New York, NY}

\begin{document}

\section{Introduction}

QCD with three or more massless quark flavors would have a first order
finite temperature transition~\cite{Pisarski:1983ms}.  With increasing
quark masses, the first order transition ends at a second order
critical point, where the correlation length of the flavor singlet
scalar meson diverges~\cite{Gavin:1993yk}.

Lattice QCD provided evidences~\cite{Aoki:1998gia,Liao:2001en} of
small singlet scalar meson screening masses, by simulations of the
naive staggered fermion action, using a temporal extend, $N_t = 4$.
Since then, no new simulations have been exploring this area of
research, partly because of the increasing difficulties of locating
first order thermal transitions with improved staggered fermion
formulations.

Simulating 3-flavor finite temperature lattices using the
nonperturbatively improved clover fermion action and the Iwasaki gauge
action, we are able to locate the first order transition with lattice
sizes of $24^3\times8$ and $32^3\times8$.  This gives us the
opportunity of studying the screening masses near the critical
endpoint on the first order transition line.

\section{Measurement techniques}

We measure the connected screening propagators on three spatial
directions, using multiple point sources either with or without
Gaussian smearing, and point sinks.  We use Gaussian smearing
parameters, $r = 4$ and $n = 100$, so the source vector centered at
$x_0$ is
\begin{equation}
\phi(x) = \left(1 + \frac{r^2}{4n}\nabla^2\right)^n \delta(x-x_0),
\end{equation}
where
\begin{equation}
\nabla^2\chi(x) = \sum_{i=1,2,3}
U_i(x)\chi(x+\hat{i}) + U_i^\dag(x-\hat{i})\chi(x-\hat{i}) - 2\chi(x).
\end{equation}

The flavor singlet screening correlator after zero momentum projection
for $N_f$ quark flavors,
\begin{align}
  \big\langle
  \sum_q[\bar{q}(x)q(x)]\sum_q[\bar{q}(y)q(y)]
  \big\rangle
  &= N_f^2 \pbp^2
    + A (e^{-m|x-y|} + e^{-m|x-y-N_l|})
    + \cdots \\
  &= - N_f \big\langle\tr[S(x,y)S^\dag(y,x)]\big\rangle
    + N_f^2 \big\langle\tr[S(x,x)] \tr[S(y,y)]\big\rangle,
\end{align}
has a nonzero vacuum state due to finite chiral condensate, and
requires evaluating disconnected diagrams.  We call the exponential
decay part the screening propagator, and write it as
\begin{align}
  e^{-m|x|} + e^{-m|N_l-x|}
  &\propto -C(x) + N_f D(x), \\
  C(x)
  &= \big\langle\tr[S(x_0,x_0+x)S^\dag(x_0+x,x_0)]\big\rangle, \\
  D(x)
  &= \big\langle\tr[S(x_0,x_0)] \tr[S(x_0+x,x_0+x)]\big\rangle
    - \big\langle\tr[S(x_0,x_0)]\big\rangle^2 \\
  & = \Big\langle
    \big(
    \tr[S(x_0,x_0)] - \pbp
    \big)^2\Big\rangle
    - \Big\langle\frac{1}{2}
    \big(
    \tr[S(x_0,x_0)] - \tr[S(x_0+x,x_0+x)]
    \big)^2\Big\rangle\label{eq:dx_sep}
\end{align}
where averaging over $x_0$ is implicit for brevity, $N_l$ is the
spatial extent of the lattice, $S$ is the unsmeared quark propagator
with $\big\langle\tr[S(x_0,x_0)]\big\rangle=\pbp$, and $C$ and $D$ are
the connected and disconnected parts of the propagator respectively.
The separation of the two terms in Eq.~\ref{eq:dx_sep} makes them
sensitive to the statistical noise from evaluating $\tr[S(x,x)]$.
They help us determine the required statistics and the computational
cost for our method detailed below.

We use complex $Z(2)_\Re \times Z(2)_\Im$ volume random sources,
applying the idea of probing\textemdash{}instead of diluted uniform
sources in the reference~\cite{Tang2012Probing}, we use stochastic
sources with space-time dilution using the greedy multi-coloring
algorithm.  We test a few configurations with different values of
$\dmin$, which is the minimum number of links between nonzero sites in
a diluted vector.  We also combine it with spin-color
separation~\cite{Wilcox:1999ab}, for the extra benefit verified in our
tests.  Table~\ref{tab:dmin-improvement} shows a representative test
result, for a particular gauge configuration of a lattice size
$32^3\times8$.  While increasing $\dmin$, the number of space-time
diluted vectors increases exponentially with the exception of
$\dmin = 2$, $4$, and $8$, which give relatively small number of
diluted vectors due to symmetry.  For $\dmin=1$, which means no
space-time dilution, spin-color separation alone does not help in
reducing the standard deviation for the same cost.  Using spin-color
separation and space-time dilution together, we find increasingly
smaller standard deviations with increasing $\dmin$ at the same
computational cost.

\begin{table}
  \centering
  \begin{tabular}{c|rrrrrrrr}
    $\dmin$                      & $1$  & $2$  & $3$  & $4$  & $5$   & $6$   & $7$   & $8$   \\
    No. of diluted vectors       & $1$  & $2$  & $23$ & $16$ & $120$ & $210$ & $411$ & $256$ \\
    $\sigma_{\text{imp}}/\sigma$ & 1.14 & 0.82 &      & 0.58 &       &       &       & 0.28  \\
  \end{tabular}
  \caption{\label{tab:dmin-improvement}Number of space-time diluted
    vectors for a $32^3\times8$ lattice corresponding to different
    values of $\dmin$, and the resulting improvement ratio of
    standard deviations.  $\sigma_{\text{imp}}$ is the standard
    deviation of $\pbp$ computed with spin-color separation and
    space-time dilution; $\sigma$ is from spin-color combined
    undiluted stochastic sources.  Standard deviations are estimated
    at the same computational cost at a fixed CG iteration number of 25.}
\end{table}

The improvement factor of four (with $\dmin=8$ in
table~\ref{tab:dmin-improvement}) comes if we solve $256\times12$
vectors for one configuration, which is impractical.  The truncated
solver method (TSM)~\cite{Collins:2007mh} brings us further
improvements when we extend it to use multiple levels of conjugate
gradient (CG) truncations and apply spin-color separation and
space-time dilution with different $\dmin$ at different truncation
levels hierarchically.  This is our method of hierarchical truncations
with stochastic probing (HTwSP).

One possible caveat: at a fixed CG iteration before converging, given
the same initial guess, space-time dilutions with different $\dmin$
produce different solutions.  Combining results from different $\dmin$
at different CG truncation levels would potentially contain bias.

\begin{figure}
  \centering
  \subfloat[][Results at the eighth CG iteration]{\includegraphics[width=.48\textwidth]{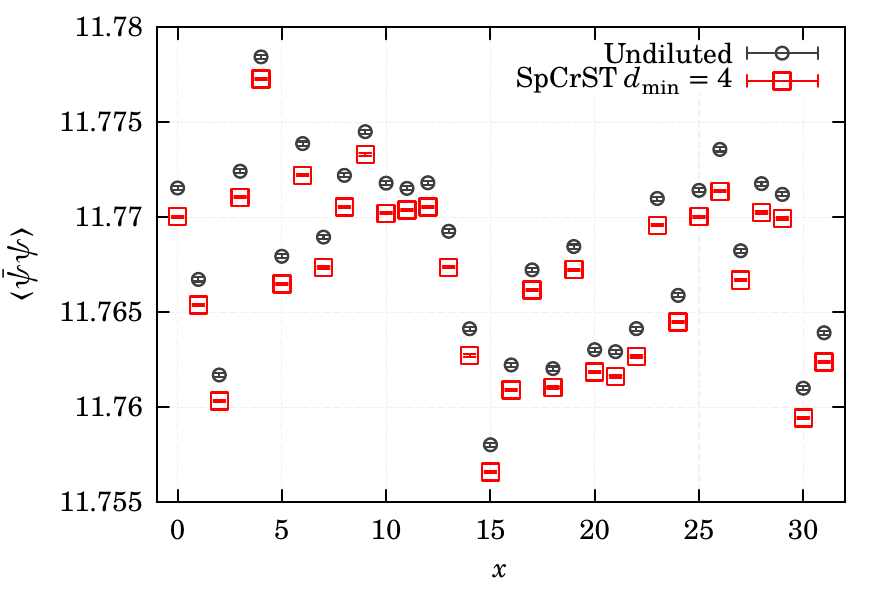}}
  \quad
  \subfloat[][Difference as a function of CG iteration]{\includegraphics[width=.48\textwidth]{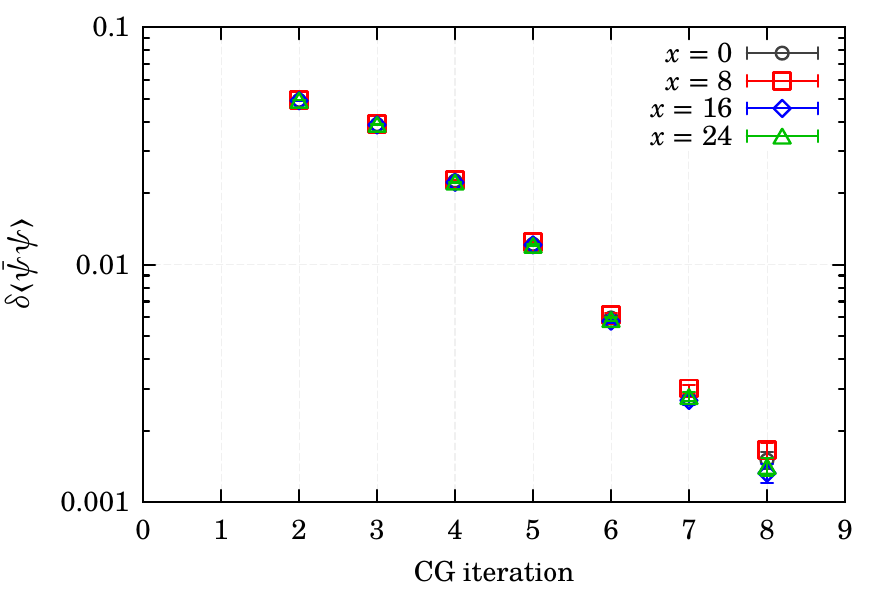}}
  \caption{\label{fig:compare-dilution}Results from the first eight CG
    iterations using undiluted random sources and diluted random
    sources of spin-color separation and space-time dilution with
    $\dmin=4$.}
\end{figure}

To test how big the difference is, we compare the undiluted results
from $49152$ volume random sources, and the diluted results from 256
source vectors per spin-color with space-time dilution of $\dmin = 4$,
which is also $49152$ vectors ($=256\times12\times16$).
Figure~\ref{fig:compare-dilution} shows $\langle\bar\psi\psi\rangle$
at each spatial slices for one tested configuration of an
ordered-start ensemble with a lattice size of $32^3\times8$,
$\kappa=0.14085$, and $\beta=1.73$.  The figure shows results from the
first eight CG iterations with a zero vector as the initial guess.
Results from these two schemes are different at finite CG iterations.
This difference, however, appears to decay exponentially.  At a few
hundred iterations where truncated results are used in this study,
though untested, the difference should be well below statistical
errors or even machine precision, so that no bias can be introduced in
this way.

For the HTwSP scheme in this study, we pick a handful of gauge
configurations and search for the best combinations of CG truncation
levels and values of $\dmin$ for each level, while respecting our cost
constraints and statistical precision needed for our study.
Table~\ref{tab:HTwSP-TSM-improvements} shows the best parameters we
find on one of the test configurations, for the original one level
TSM, two level and three level TSM (2L-TSM, 3L-TSM), and two level
HTwSP.  We try to measure and minimize the computed ratios of
variances, $R_{\text{imp}} = \sigma^2_{\text{imp}}/\sigma^2$, improved
against the undiluted one.  For a fixed cost equivalent to about $500$
convergent CG inversions, we find more than $40$ times speedup using
the scenario of 2L-HTwSP.  Further speedup is guaranteed with a larger
cost constraint.

\begin{table}
  \centering
  \begin{tabular}{lrrrr}
    Scenario                    & TSM      & 2L-TSM   & 3L-TSM   & 2L-HTwSP  \\
    Cost                        & 500.3    & 493.2    & 486.2    & 501.9     \\
    \hline
    \(R_{\text{imp}}\)          & 0.203(4) & 0.119(4) & 0.102(3) & 0.0231(6) \\
    \hline
    \(N_{\text{h}}\)            & 99       & 21       & 5        & 1         \\
    \(\text{Iter}_{\text{h}}\)  & 2049     & 2049     & 2049     & 2049      \\
    \(C_{\text{h}}\)            & 1        & 1        & 1        & 24        \\
    \hline
    \(N_{\text{l}}\)            & 2990     & 1312     & 442      & 4         \\
    \(\text{Iter}_{\text{l}}\)  & 275      & 425      & 475      & 475       \\
    \(C_{\text{l}}\)            & 1        & 1        & 1        & 192       \\
    \hline
    \(N_{\text{l2}}\)           &          & 8200     & 2964     & 1         \\
    \(\text{Iter}_{\text{l2}}\) &          & 50       & 150      & 200       \\
    \(C_{\text{l2}}\)           &          & 1        & 1        & 3072      \\
    \hline
    \(N_{\text{l3}}\)           &          &          & 13255    &           \\
    \(\text{Iter}_{\text{l3}}\) &          &          & 25       &           \\
    \(C_{\text{l3}}\)           &          &          & 1        &           \\
  \end{tabular}
  \caption{\label{tab:HTwSP-TSM-improvements}Improvements of different
    number of truncation levels, and one particular HTwSP scheme.
    Parameters are constrained by the computational cost.  A `Cost' of
    $1$ equals computational cost of one convergent CG inversion.
    $R$ is the ratio of variances (improved against undiluted one),
    which equals the inverse of the speedup in the wall-clock time for
    a fixed statistical error.  While the subscripts denote different
    truncation levels, parameters for each level are: $N$ is the
    number of random vector sets, the number of undiluted stochastic
    sources; $C$ is the number of random vectors in one set, counting
    in dilution cost; `Iter' is the number of fixed CG iterations on
    that truncation level (2049 is the convergent iteration number).}
\end{table}

\section{Simulation results}

\begin{table}
  \centering
  \begin{tabular}{rrrlrlll}
    $\beta$ & $\kappa$ & $N_l$ & start & $N_{\text{conf}}$ & plaquette    & gauge        & polyakov     \\
    \hline
    1.73    & 0.14085  & 24    & ord   & 2930              & 0.528741(67) & 1.24940(19)  & 0.00831(16)  \\
    1.73    & 0.14085  & 32    & ord   & 13875             & 0.528701(34) & 1.249514(98) & 0.008315(56) \\
    \hline
    1.73    & 0.14085  & 24    & dis   & 4060              & 0.51495(15)  & 1.28832(43)  & 0.002555(73) \\
    1.73    & 0.14085  & 32    & dis   & 16815             & 0.514765(55) & 1.28883(16)  & 0.002480(22) \\
    \hline
    1.74    & 0.14054  & 32    & ord   & 16000             & 0.529832(44) & 1.24603(13)  & 0.009558(32) \\
    \hline
    1.74    & 0.14054  & 32    & dis   & 10450             & 0.520610(98) & 1.27195(28)  & 0.003606(65) \\
  \end{tabular}
  \caption{\label{parameter-stats}The expectation values of plaquette,
    gauge density, and Polyakov loop.  Number of configurations has
    first 1000 removed as thermalization.}
\end{table}

We start simulations with ordered and disordered gauge configurations
with lattice sizes of $24^3\times8$ and $32^3\times8$.  For our two
lattice coupling values of $\beta = 1.73$ and $1.74$, we have tested a
few $\kappa$ values and located the two-states signal at
$\kappa = 0.14085$ and $0.14054$ respectively.
Table~\ref{parameter-stats} shows simple ensemble statistics.  We
measure screening propagators every 20 trajectories.

\begin{figure}
  \centering
  \subfloat[][$\beta=1.73$, $\kappa=0.14085$]{\includegraphics[width=.48\textwidth]{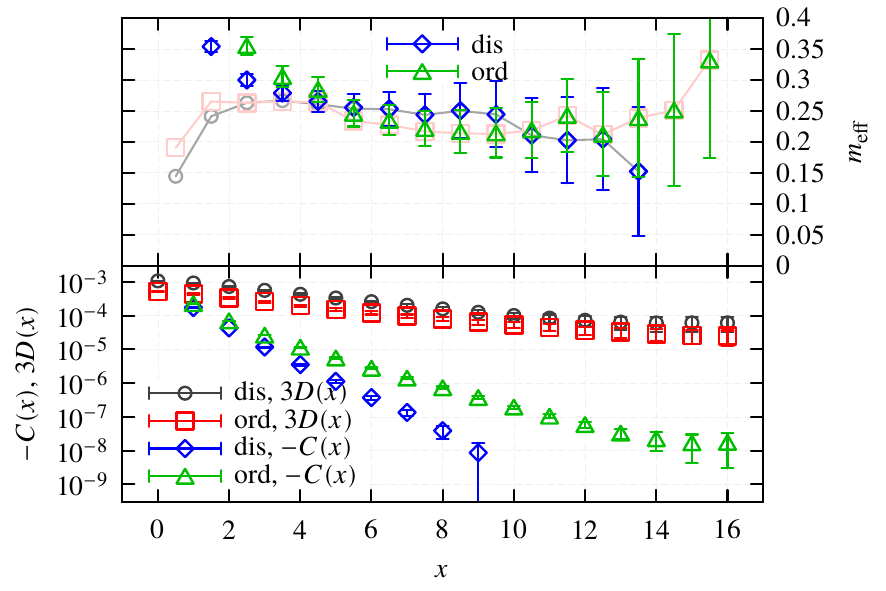}}
  \quad
  \subfloat[][$\beta=1.74$, $\kappa=0.14054$]{\includegraphics[width=.48\textwidth]{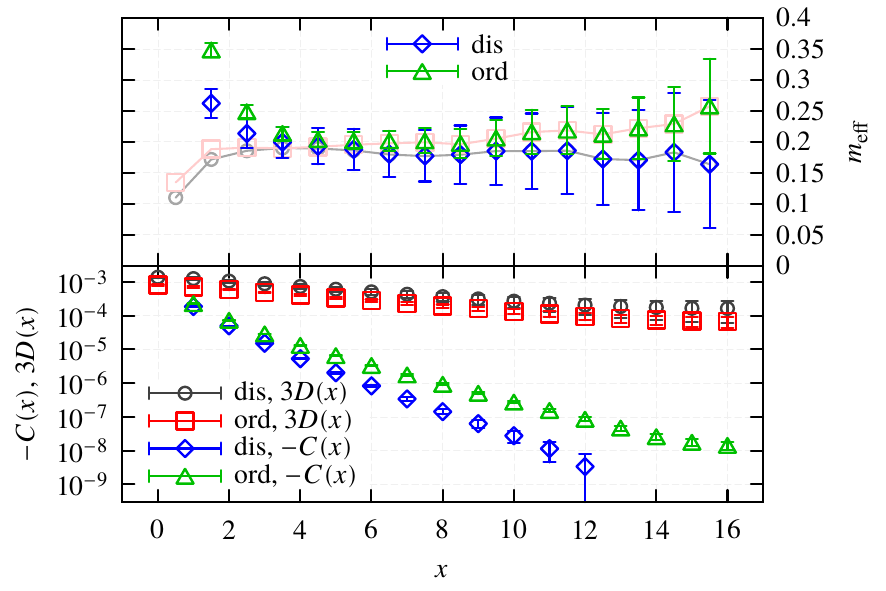}}
  \caption{\label{fig:prop_effm_SC}Lower panels show the connected and
    disconnected screening propagators of the flavor singlet scalar
    meson.  Upper panels show the effective mass obtained from the
    full propagators, and the ones from disconnected propagators
    without error bars using light color as comparison.  In the keys,
    `dis' stands for the disordered-start ensemble, and `ord' stands
    for the ordered-start ensemble.}
\end{figure}

Figure~\ref{fig:prop_effm_SC} shows the connected and disconnected
parts of the flavor singlet scalar meson propagator and its effective
masses as a solution to propagator values at nearby two spatial
separations.  Because the disconnected part are orders of magnitude
larger than the connected part, at long spatial separations, the
effective masses from the full singlet propagator and the effective
masses from the disconnected part alone agree.  We see no statistical
difference between the screening masses from disordered-start and
ordered-start ensembles, which respectively correspond to the zero
temperature and the finite temperature phases.

\begin{figure}
  \centering
  \subfloat[][$\beta=1.73$, $\kappa=0.14085$]{\includegraphics[width=.48\textwidth]{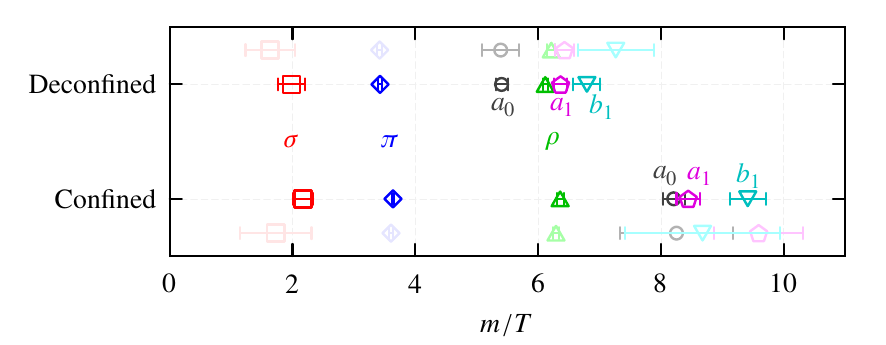}}
  \quad
  \subfloat[][$\beta=1.74$, $\kappa=0.14054$]{\includegraphics[width=.48\textwidth]{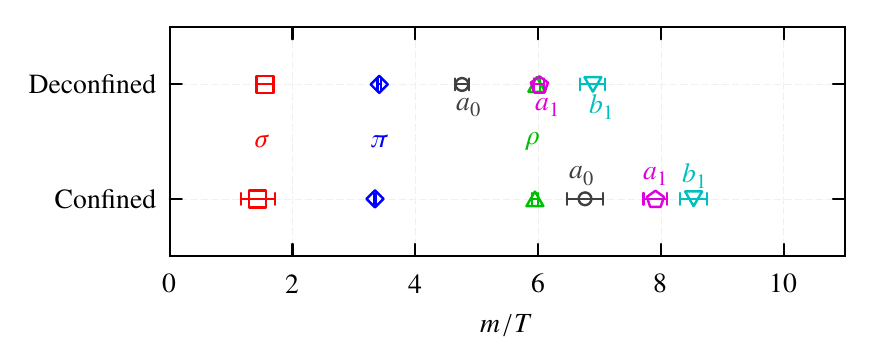}}
  \caption{\label{fig:screening_mass_hier}Screening mass hierarchy
    across the first order transition.  Symbols with light color
    represent results from $24^3\times8$ lattices, while the rest are
    from $32^3\times8$ lattices.}
\end{figure}

Figure~\ref{fig:screening_mass_hier} shows meson screening masses at
the transition.  The flavor singlet scalar $\sigma$ meson is about
half as light as the pion, and continues to decrease as the system
approaches the critical endpoint, from $\beta=1.73$ to $\beta=1.74$.
In the two parameter sets we have, the screening masses of the $\rho$,
the $\pi$, and the $\sigma$ mesons are almost constant across the
transition.  We see the pion screening mass is smaller in the
deconfined phase than the confined phase at $\beta=1.73$,
$\kappa=0.14085$, while it is in an opposite order in the other
parameter set.  This is different from previous
results~\cite{Aoki:1998gia,Liao:2001en} with staggered fermions.  As
the system undergoes the first order transition from confined zero
temperature phase to deconfined finite temperature phase, the $a_1$
meson becomes degenerate with the $\rho$ meson, and the $a_0$ meson
drops and becomes closer to the $\pi$ meson, as expected.

\begin{figure}
  \centering
  \includegraphics{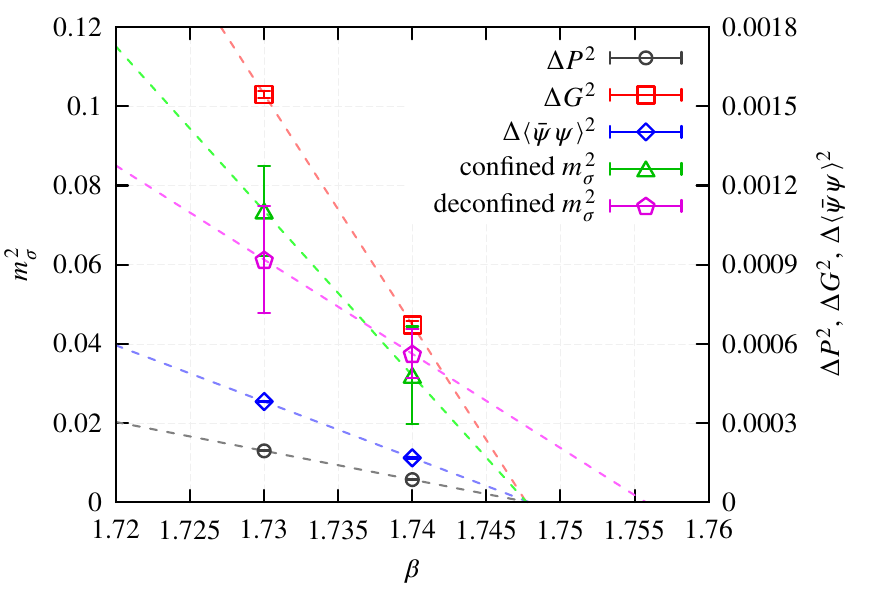}
  \caption{\label{fig:squared_linear}Linear extrapolations of
    $m_\sigma^2$ and the squared discontinuities of plaquette, gauge
    density, and $\langle\bar\psi\psi\rangle$.}
\end{figure}

Near the second order critical point, the system should exhibit the
behavior of critical scalings.  The flavor singlet scalar meson
screening masses should scale as the inverse of the correlation
length, $m_\sigma \propto t^\nu$, as proposed by Gavin \textit{et
  al.}~\cite{Gavin:1993yk}.  Other observables, we expect, should
scale as a combination of the magnetic like order parameter,
$\propto t^{\bar\beta}$ (we write it as $\bar\beta$ to avoid confusion
with the lattice coupling), and the energy like quantities,
$\propto t^{-\alpha}$.  Since we lack enough data sets to
differentiate different universality classes, we treat the data as the
mean field theory, such that $\alpha = 0$ and $\bar\beta = \nu = 1/2$.
With the choice of the reduced temperature
$t = (\beta - \beta_{\text{C}})/\beta_{\text{C}}$, where
$\beta_{\text{C}}$ is the lattice coupling at the critical point, if
the system is sufficiently close to the critical point, $m_\sigma^2$
should be linear in $\beta$, and the squared discontinuities of the
plaquette, the gauge density, and the chiral condensate
$\langle\bar\psi\psi\rangle$, should also be linear in $\beta$.
Figure~\ref{fig:squared_linear} shows these squared quantities, with
linear lines going through two data points of each observable.  Given
the noisy scalar singlet meson screening masses, our extrapolations
give consistent $\beta_{\text{C}}$.  The mean field theory thus
predicts that the critical endpoint is between $\beta=1.745$ and
$1.75$, which is consistent with our results from analysis of kurtosis
intersections~\cite{Jin:2014hea}.  However, we need more data sets to
determine the correct universality class.

\section{Summary}

We locate the first order transition with two-states signal at two
parameter sets using the nonperturbatively improved clover fermion
action and the Iwasaki gauge action.

We developed an easy-to-implement method of hierarchical truncations
with stochastic probing, which gives us an estimated $40\times$
speedup in measuring the quark condensate at spatial lattice slices.

Our data show the $\sigma$ screening mass is about half of the pion
mass on both sides of the first order transition close to the
endpoint.  A preliminary extrapolation to the critical endpoint shows
consistency between our data and the mean field critical scaling.  To
determine the correct universality class requires more statistics and
expanded parameter sets.

\acknowledgments

The numerical calculations have been done on K computer at RIKEN AICS.
Our lattice evolutions uses BQCD, while measurements use my modified
code shared by the lattice group at the University of Helsinki.

This work is supported in part by the Grants-in-Aid for Scientific
Research from the Ministry of Education, Culture, Sports, Science and
Technology (No.~26800130).

\bibliographystyle{JHEP}
\bibliography{ref}

\end{document}